\input jytex
\input epsf
\hsize=6.5in   	\vsize=9in
\topmargin=1in	\leftmargin=1in
\typesize=12pt
\baselinestretch=1300
\pagenumstyle{arabic}
\sectionnumstyle{arabic}
\newcount\fignum  \fignum=0
\def\fignumstyle#1{\global\expandafter\let\expandafter
	\fignumtype\csname#1\endcsname}
\def\figlabel#1{\global\advance\fignum by1
  \label{#1}{\hbox{\fignumtype\fignum}}\putlab{#1}}
\def\Figure#1{\hbox{Fig.~\figlabel{#1}}}

\def\putfig#1#2#3{ \vbox{ \bigskip\medskip
	\centertext{\epsffile{#2}}
  \centertext{ {\bf Figure \putlab{#1}.} #3} \bigskip}}

\fignumstyle{arabic}

\def\section#1{\bigskip\goodbreak\newsectionnum=\next
	\lefttext{\bf\arabic\sectionnum. #1}\nobreak\medskip}
\def\subsection#1{\bigskip\goodbreak
	\lefttext{\it\arabic\sectionnum.#1}\nobreak\medskip}
\def\appendix#1{\bigskip\goodbreak\newsectionnum=\next
	\lefttext{\bf Appendix \Alphabetic\sectionnum. #1}\nobreak\medskip}
\def\eq#1{\eqno\eqnlabel{#1}}
\def\ref#1{\markup{[\putref{#1}]}}


\def\TITLE#1{\vskip .5in \centertext{\bigsize #1}}
\def\AUTHOR#1{\vskip .3in\centertext{#1}}
\def\ANDAUTHOR#1{\vskip .1in\centertext{\rm and}
	\vskip .1in\centertext{#1}}
\def\ABSTRACT#1{\vskip .5in \vfil\centertext{\normalsize\bf Abstract}
	\smallskip #1 \vfil}

\def\disclaimer{
\vskip1cm\hphantom{VVVV}

\vfill
\centertext{\bf Disclaimer}
\vskip1cm

\begin{narrow}
{\smallsize
This document was prepared as an account of work supported
by the United States Government. Neither the United States Government nor
any agency thereof, nor The Regents of  the University of California, nor
any of their employees, makes any warranty, express or implied, or assumes
any legal liability or responsibility for the accuracy, completeness, or
usefulness of any information, apparaus, product, or process disclosed, or
represents that its use would not infringe privately owned rights. Reference
herein to any specified commercial products procerss, or service by its
trade name, trademark, manufacturer, or otherwise, does not necessarily
constitute or imply its endorsement, recommendation, or favoring by the
United States Government or any agency thereof, or the Regents of the
University of California. The views and opinions of authors expressed herein
do not necessarily state or reflect those of the United States Government 
or any agency thereof, or the Regents of the University of California
and shall not be used for advertising or product endorsement purposes.}
\end{narrow}
\vfill
\centerline{\it Lawrence Berkeley Laboratory is an equal opportunity
employer.}
\vfill\eject}
\def\bra{\langle}
\def\ket{\rangle}
\def\lb{\left[}
\def\rb{\right]}

\def\co{{\cal O}}
\def\cp{{\cal P}}
\def\cq{{\cal Q}}

\def\Res{\mathop{\rm Res}}
\def\gh{\,{\rm gh}\,}
\def\dim{\,{\rm dim}\,}
\def\mod{\,{\rm mod}\,}
\def\ord{\,{\rm ord}\,}
\def\ie{{\it i.e.}}
\def\etal{{\it et al.}}

\def\sumdel{\sum_\Delta}
\def\ldb{\lbrack\!\lbrack}
\def\rdb{\rbrack\!\rbrack}
\def\lll{\lbrace\!\!\lbrace}
\def\rrr{\rbrace\!\!\rbrace}


\begin{titlepage}
{\hbox to\hsize{\footnotesize\baselineskip=12pt
		\hfil\vtop{\hbox{\strut UCB-PTH-92/30}
		\hbox{\strut LBL-32850}
		\hbox{\strut CALT-68-1788}
		\hbox{\strut hep-th/9210106}}}}

\TITLE{Solving Topological 2D Quantum Gravity Using Ward Identities}

\AUTHOR{{\sc David Montano}\footnote[1]{E-mail:\tt\
montano@theory3.caltech.edu}}
\smallskip
\centertext{\it California Institute of Technology, Pasadena, CA 91125}

\ANDAUTHOR{{\sc Gil Rivlis}\footnote[2]{E-mail:\tt\
gil@physics.berkeley.edu}}
\smallskip
\centertext{\it Department of Physics, University of California at
Berkeley\\and\\Theoretical Physics Group, Lawrence Berkeley
Laboratory\\Berkeley, CA 94720}

\ABSTRACT{
A topological procedure for computing correlation functions for {\it any}
$(1,q)$ model is presented. Our procedure can be used to compute any 
correlation function on the sphere as well as some correlation functions 
at higher genus. We derive new and simpler recursion relations that extend
previously known results based on $W$ constraints. In addition, we compute
an effective contact algebra with multiple contacts that extends Verlindes'
algebra. Computational techniques based on the KdV approach are developed 
and used to compute the same correlation functions. A simple and elegant 
proof of the puncture equation derived directly from the KdV equations is 
included. We hope that this approach can lead to a deeper understanding of 
$D=1$ quantum gravity and non-critical string theory.}
\vfill
\lefttext{October, 1992}

\end{titlepage}

\pagenumstyle{roman}

\disclaimer

\pagenumstyle{arabic}

\pagenum=0	


\section{Introduction}

In recent years a class of models of two-dimensional gravity has
been solved exactly. These models could be 
variously interpreted as $(p,q)$
conformal field theories coupled to two-dimensional gravity (\ie,
Liouville field) or the continuum limit of a spin system on a random
lattice. These interpretations correspond to the KdV and matrix model
approaches, respectively. It was hoped that these solutions would
provide insight into higher dimensional quantum gravity or non-critical
string theory. 
Though these models are in principle exactly solvable, there were
technical complications in providing explicit general solutions for
correlation functions; \ie, solving the non-linear differential
equations of the KdV hierarchy. Thus, it was difficult to develop
intuition about the solutions. In this paper we provide a simple
geometrical method for computing correlation functions for the 
$(1,q)$ models. Any correlation function on the sphere can be computed;
partial results for higher genus will also be presented.
The advantage of our method over that
of the $W$-constraints is that our constraints (or Ward identities)
can be written instantly for theories involving any number of primary
fields. $W$-constraints are not known explicitly for anything greater
than $W_3$.

It has been suggested that the partition function for 
$c<1$ quantum gravity is given by the square of a $\tau$ function
of the KdV hierarchy satisfying the string equation. This followed
the work on the matrix model formulation of two dimensional 
gravity\ref{brez,bess,kaza2,davi,kaza3,kaza4} and its double scaling 
limit\ref{brez2,doug2,gros}. 
Douglas suggested that the KdV hierarchy along with the string 
equation may be used to describe minimal matter coupled to two
dimensional gravity\ref{doug3}. The work of Dijkgraaf 
\etal\ref{dijk3} and Fukuma \etal\ref{fuku}
proved that the $\tau$ function for 2D gravity satisfies a set
of constraints given by generators of the Virasoro algebra. These
generators act on the 
infinite dimensional space of all flow parameters of the KdV
hierarchy. The constraints were used to derive recursion
relations for the $(1,q)$ models of 2D gravity. These
models are the critical
topological points of minimal matter coupled to 2D gravity. For the
$(1,2)$ models, which have only one primary field, the Virasoro constraints
are sufficient to completely solve the model. They are entirely
consistent with the KdV approach.

The Virasoro constraints are not sufficient in themselves to completely
solve the models with more than one primary field. It was then suggested
that since the only set of operators forming a closed algebra with
the Virasoro constraints are the $W_n$ generators, perhaps
the recursion relations necessary to
completely solve the $(1,q)$ models could be derived by imposing
the $W_n$ constraints. Goeree\ref{goer}  has derived the $W_3$ 
constraints from the string equation and obtained the corresponding
recursion relations.
These recursion relations are difficult to derive and provide
very little intuition into the structure of quantum gravity. To
obtain explicit recursion relations for models with more than 
two primary fields using $W_n$ constraints is an extremely
tedious process which has not yet been done. 

Shortly after the  matrix model and double scaling limit breakthrough, an
alternative approach was suggested by Witten based on topological
field theory ideas\ref{witt2}. Subsequently, the so-called topological gravity, 
was developed further and it was shown to be related to
the `regular' version in a direct way\ref{dist,verl2,dijk2,eguc,li,dijk4}. 

We will present a very simple and intuitive
procedure for computing any correlation function (including
that of primary fields) in topological gravity with an arbitrary
number of primary fields. Some higher genus results are also derived,
including the one point function on the torus. This is a very interesting
result because it is not computable from the $W_n$ constraints. 

Our procedure involves imposing a Ward identity of a clearly geometrical
origin (though we cannot derive it from a field theory). This Ward
identity is superficially similar to the $W$-constraints.
It involves the  evaluation of a correlation function by summing
over all possible surface degenerations. At each degeneration, a ``complete''
set of states is inserted. Contact terms between the operators are 
then evaluated at the degeneration point. Multiple contacts
and degenerations occur for the $(1,q)$ models (for $q>2$).
We will show that the actual numerical values for the contact terms
are irrelevant. All that matters is the number of contacts (that depends
on the primary field ``doing'' the contact),
ghost number conservation and one- and two-point correlation functions.
(From the two-point function (``metric'') an ``identity''
operator will be constructed.)
In order to achieve this we had to introduce ``anti-states'' or
operators of negative dimension. These are, in our interpretation,
unphysical operators whose higher point correlation functions (greater 
than two)
vanish. This will be formulated more precisely in the body of the paper.

Our method relies on intuition from both  topological and  KdV
approaches to quantum gravity.
Some techniques for arriving at
certain exact results directly from the KdV approach
are presented in the appendices. We review our
derivation of ghost number conservation for the $(1,q)$ models from the
KdV and derive the metric used in the body of the paper. Perhaps, one of
the more elegant results is a simple proof of the puncture equation for
topological gravity directly from the KdV equations.

This paper is then organized as follows. In Section 2 we introduce
the anti-states and discuss the ``metric'' and the ``identity operator''.
In section 3 we present our method on the sphere
and derive the correspondence with previous known results.
In Section 4 we compute effective contact terms and derive
the Virasoro and $W_3$ constraints.
In Section 5 we investigate the degeneration equation for
the torus and higher genus. In the conclusion open questions and
directions for future research are discussed. A few appendices
are included. In Appendix A we review the KdV approach to quantum gravity,
in Appendix B we review the $(1,q)$ models and the ghost number 
conservation rule. Appendix C is devoted to the extension of the 
KdV approach to anti-states and the derivation of the metric. In
Appendix D we prove the puncture equation from the KdV and
in appendix E we review the $W_3$ constraints.

\section{Anti-States}

In order to describe the Ward identities (or recursion relations)
systematically, we first need to introduce the concept of an
``anti-state'' and identity operator. By anti-states we mean the
states of negative dimension which are conjugate to the physical
states on the sphere. These are necessary in order to have a 
non-vanishing two-point
function or metric. Typically, of course, in a field theory the 
anti-states are also
physical states. The appropriate analogy is that of momentum and
position states. In quantum gravity there is a difficulty: the anti-states
are not physical. Two-point functions of physical states vanish.
Nevertheless, we will introduce these conjugate states in order
to make sense of the theory. They will be unphysical states whose
higher point functions (greater than two) will be set to zero. This
will then allow for an elegant prescription for deriving a complete
set of Ward identities that will determine all of the correlation
functions of the $(1,q)$ models of two-dimensional gravity.

Our
procedure has many features reminiscent of a field
theory interpretation. It is also analogous with the $W$-constraints.  
The results are in complete agreement with the
KdV equations. Following the Verlindes\ref{verl2}
 we attempted to determine recursion
relations from the possible contact terms and surface degenerations.
We found to our surprise that if we included only all possible surface
degenerations, we could reproduce the Virasoro constraints and recursion
relations for the other primary fields consistent with the KdV. At each
surface degeneration we would insert a ``complete'' set of states which would
come in contact with the operator at the degeneration. Contact terms would
arise effectively whenever one of the degenerating surfaces corresponded
to a trivial correlation function. This procedure proved general enough
to compute any correlation function on the sphere, including those
of primary fields. Partial results for higher genus are also computable.

\subsection{1. Defining the Hilbert Space}

We will now discuss the basic structure of the correlation
functions of the $(1,q)$ models.\footnote[*]{The $(1,q)$ models are defined in
Appendix B.}
 Our construction will be based in analogy
to standard quantum mechanics (without gravity) where one can define a Hilbert
space with an adjoint, metric, and identity operator such
that correlation functions can be evaluated as follows,
$$\bra\alpha\vert\beta\ket=\sum_{n,m}\bra\alpha\vert m\ket \eta^{mn}\bra n
\vert\beta\ket, \eq{fact}$$
where $\vert n\ket$ are a complete set of states such that
$I=\sum_{m,n}\vert m\ket\eta^{mn}\bra n\vert$
is the identity and $\bra m\vert n\ket=\eta_{mn}$ is the invertible metric.
In quantum gravity there are difficulties in defining a meaningful Hilbert
space. In the
$(1,q)$ models of two-dimensional gravity 
it turns out that if one chooses the
two-point function to be the metric, then the  adjoint is not
in the spectrum of original states. Indeed, all two point functions vanish.
Nevertheless, one can define a non-vanishing two-point function by
extending the set of allowed states.
These new unphysical states
will appear only on the boundaries of moduli space and
simulate  the interaction of the physical states with the boundaries. They
never appear in the final correlation functions. In the following
we will show precisely how this procedure works. But first we will develop
the necessary tools for evaluating the correlation functions.

\subsection{2. The Metric on the Boundary}

From the KdV formulation of two-dimensional gravity one can show that the
two-point function on the sphere for the $(1,q)$ models is given by 
(C.17), (see Appendix C)
$$\bra \cp_i\cp_j \ket_0 = |i|\delta_{i+j}.\eq{2point}$$
Since the operators, $\cp_i$, exist only for $i>0$, we see that the two-point
function always vanishes for physical operators. Thus, we seem to be
unable to write a metric. Let us circumvent this problem by
going ahead and defining a formal
 adjoint operation as follows,
$$\cp_i^{\dagger}=\cp_{-i},\eq{adjoint}$$
and, thus, extend the space of operators to include $\cp_i$ with
$i<0$. These operators have negative dimension.
 We now have an invertible metric defined by the two point function,
$$\bra \cp_i\cp_j\ket_0=\eta_{ij},\eq{metric1}$$
where $\eta_{ij}=|i|\delta_{i+j}$. We still have to find an identity operator
to complete our Hilbert space. We also need to show that the operators,
$\cp_i$ with $i<0$, decouple (except for the metric). 

\subsection{3. Definition of the Identity Operator}

We must now define an identity operator. This is a more subtle problem.
The naive choice,
$$I=\sum_{j,i
\neq 0(\mod q)} \vert \cp_i\ket \eta^{ij} \bra \cp_j\vert,\eq{identity}$$
where $\eta^{ij}$ is the inverted metric (``propagator''),
does not reproduce the correct structure of the $(1,q)$ gravity. In some 
sense this problem cannot be resolved. An identity operator cannot be
defined globally in a theory of gravity. But an identity
as in (\puteqn{identity}) can be defined ``locally'' which is, in fact,
all that we need. Because the theory is topological what we mean by local
is a degeneration of the surface marked by a point. Thus, in computing
correlation functions, (\puteqn{identity}) must be inserted at every
possible degeneration of the surface consistent with the compactification
of the moduli space. These degenerations
correspond to two points coming in contact to pinch off a sphere or a
handle (of a higher genus surface). In general, $(1,q)$ models permit
multiple degenerations (up to $q$ contacts). Then, at each degeneration
one inserts (\puteqn{identity}) and evaluates the contact terms.
The precise procedure will be explained in the next section. 

\subsection{4. The One-Point and other Correlations with Anti-States}

Although introducing negative dimension states enabled us to write a
non-vanishing two-point function, one has to check the consequences
for other correlation functions. Indeed, the one-point function on the 
sphere can be non-vanishing only if the operator is 
$\cp_{-q-1}$\footnote[*]{This is a consequence of ghost number conservation
rule. This was derived in [\putref{mont,rivl2}] and is reviewed in
the appendices.}. Thus, in a regular (\ie, without anti-states) theory,
{\it all} one-point functions vanish on the sphere. In the extended
model, however, this is not necessarily true. Since the operator
$\cp_{-q-1}$ exists, we cannot set the this one-point function to zero
arbitrarily. One has to make sure that the theory remains consistent.

From the KdV approach, one can compute this correlation and get
$$\bra \cp_{-q-1} \ket_0 = -q.  \eq{1point}$$
We see that, indeed, this function is {\it not} zero and, as we
will see later, is very important in the construction of the
degeneration equation (Ward identities).

Higher point correlation functions
with anti-states will be set to zero. This
can be done consistently. They thus never contribute to
the recursion relations. From the KdV approach we know that any correlation
function with (at least one) puncture operator and more then three
fields will always vanish if one of the fields in of negative
dimension (anti-state). It is natural to extend this to any higher
point function. A direct computation involving analytic continuation
of KdV results supports this conclusion.

\section{Computation of Correlation Functions on the Sphere}

In this section the method for evaluating correlation functions on the 
sphere will be presented. We will introduce a
diagrammatic notation that will help clarify the method.

\subsection{1. Notation and Diagrams}

We begin the computation of a correlation function by specifying an operator
at the location in which the degenerations take
place. This operator will conveniently be taken to be the first one
and will be referred to as the {\it marked} operator. \Figure{corrfunc}
represents
the correlation function $\bra \cp_n\cp_{i_1}\cdots\cp_{i_m}\ket_0$.
The marked operator will be represented by an `$\times$' in the figure.

\putfig{corrfunc}{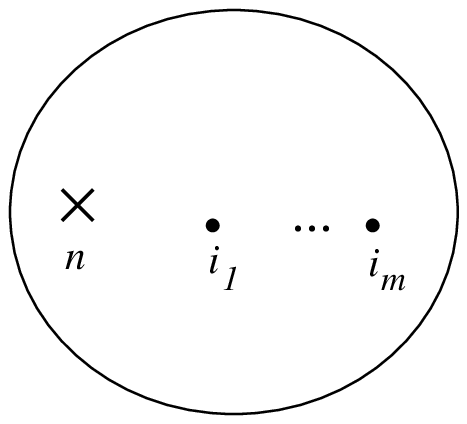}{The Correlation Function 
$\bra \cp_n\cp_{i_1}\cdots\cp_{i_m}\ket_0$.}

The marked operator $\cp_n$ is a descendant of the primary field $\cp_\alpha$,
where $\alpha = (n\mod q)$. This determines the number of contacts it
can have. Descendants of $\cp_\alpha$ will have contacts with $\alpha$
operators. We will represent a contact as an overbrace in the correlation
function and in figures it will be represented by a line connecting the marked
operator with the other operator in contact. For multiple contact, the 
overbrace will be over all the operators and in the figures we will have a line
connecting the marked opertor with {\it each} operator it comes in contact
with. For example, \Figure{contact} represents the correlation function 
$\bra\overbrace{\cp_2 \cp_i \cp_j} \cp_k \cp_l \ket_0$, where $\cp_2$ is in
contact with both $\cp_i$ and $\cp_j$.

\putfig{contact}{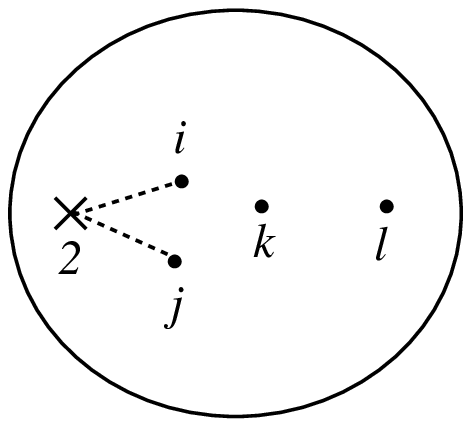}{A Correlation Function with Contacts:
$\bra\overbrace{\cp_2 \cp_i \cp_j} \cp_k \cp_l \ket_0 $.}

A degeneration of a sphere results in a split to two spheres. Multiple
degenerations result in multiple spheres. In our method, each degeneration
involves the insertion of the ``identity'' operator. In the figures we will
represent the degeneration in an obvious way, and the operators from
the identity will be subscripted with integers to distinguish them from
the original operators in the correlation function. Also, a summation will
be assumed on integer subscripts. \Figure{degen}, for
example, represent the double degeneration of $\bra \cp_a \cp_b \ket_0$ into
$\sum_{i_1,j_1,i_2,j_2} \bra \cp_a \cp_{i_1} \cp_{i_2} \ket_0 \bra \cp_b 
\cp_{j_1} \ket_0 \bra \cp_{j_2}\ket_0 \eta^{i_1j_1}\eta^{i_2j_2}$.

\putfig{degen}{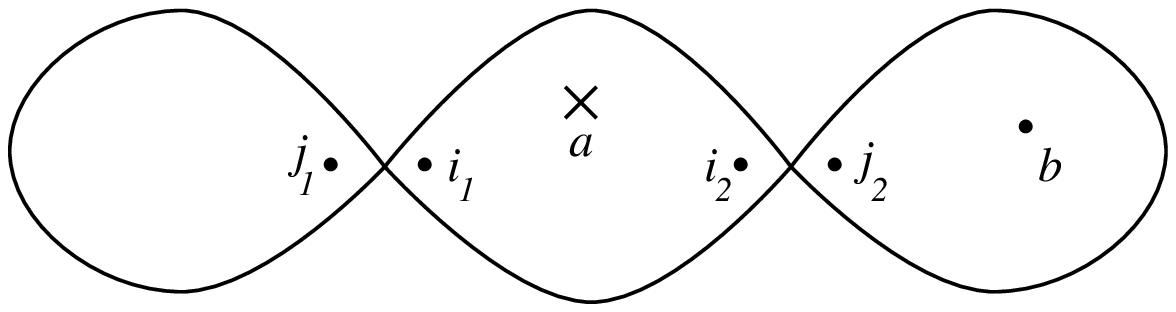}{A Double Degeneration of $\bra\cp_a\cp_b\ket$.}

The combination of degenerations and contacts is the only contribution
to correlation functions in our method. The marked operator comes in 
contact with degenerations, the number of which is determined by the 
primary field from which the marked operator descends.
In the figures, we combine the notations for contacts and degenerations
and get, for example, \Figure{degcont}. In Figure \putlab{degcont}
the operator $\cp_2$ in the correlation function $\bra \cp_2 \cp_a \cp_b\ket_0$
comes in contact with a double degeneration.

\putfig{degcont}{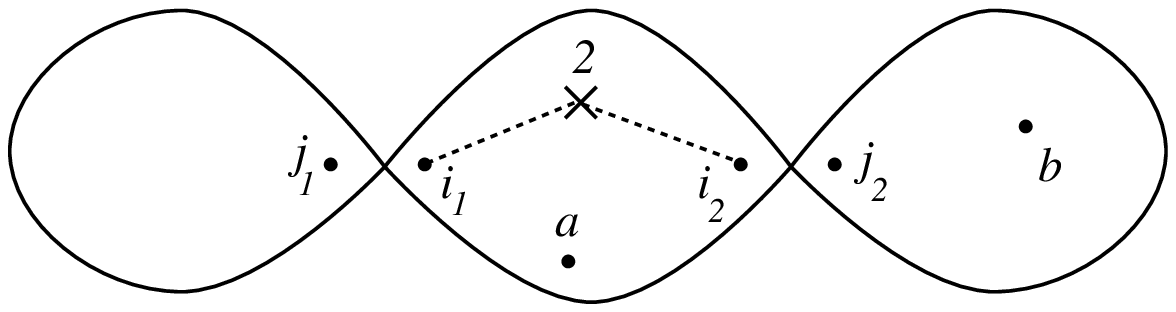}{Contact with Degeneration of 
$\bra\cp_2\cp_a\cp_b\ket_0$.}

\subsection{2. The Degeneration Equation}

We are now in a position to present the method. The main idea is the
following statement:

\begin{narrow} {\noindent \it Given a correlation function 
$\bra \cp_n \cp_{i_1}\ldots\cp_{i_m}\ket_0$, with the marked operator
$\cp_n$, then the summation over all the contacts with degenerations
of this operator vanishes.}
\end{narrow}

\noindent Formally, we will write
$$\sumdel\bra\cp_n\cp_{i_1}\ldots\cp_{i_m}\ket_0 = 0, \eq{degeq}$$
where the symbol $\sumdel$ means `sum over all degenerations when
the first operator performs the contacts'. We will call this equation
{\it the degeneration equation}.

Let us demonstrate the technique by computing a simple correlation
function. Let $q=4$ (\ie, three primary fields), and we will compute
$\bra\cp_2\cp_2\cp_1\ket_0$. $\cp_2$  is the marked operator (it is first)
and since it's a descendant of $\cp_2$ $(2\mod 4 =2)$\footnote[*]{Obviously, 
$\cp_2$ is the primary field and as such it's the $0$th descendant of itself.},
it will come in contact with two operators. Thus we have a double 
degeneration.  The degeneration equation is then
$$ \eqalign{0 = \sumdel\bra & \cp_2\cp_2\cp_1\ket_0 = \cr &
\bra\overbrace{\cp_2\cp_{i_1}\cp_{i_2}} \cp_2\cp_1 
 \ket_0	\bra\cp_{j_1} \ket_0 \bra\cp_{j_2}\ket_0\eta^{i_1j_1}\eta^{i_2j_2}+
\bra\overbrace{\cp_2\cp_{i_1}\cp_{i_2}} \cp_2\ket_0
\bra\cp_{j_1}\cp_1\ket_0\bra\cp_{j_2}\ket_0\eta^{i_1j_1}\eta^{i_2j_2} +\cr&
\bra\overbrace{\cp_2\cp_{i_1}\cp_{i_2}} \cp_2 \ket_0
\bra\cp_{j_1}\ket_0 \bra\cp_{j_2} \cp_1 \ket_0\eta^{i_1j_1}\eta^{i_2j_2}+
\bra\overbrace{\cp_2\cp_{i_1}\cp_{i_2}}\cp_1 \ket_0
\bra\cp_{j_1} \cp_2 \ket_0 \bra\cp_{j_2}\ket_0\eta^{i_1j_1}\eta^{i_2j_2}+\cr&
\bra\overbrace{\cp_2\cp_{i_1}\cp_{i_2}}\cp_1 \ket_0
\bra\cp_{j_1} \ket_0 \bra\cp_{j_2} \cp_2 \ket_0\eta^{i_1j_1}\eta^{i_2j_2}+
\bra\overbrace{\cp_2\cp_{i_1}\cp_{i_2}}\ket_0\bra
\cp_{j_1}\cp_2 \ket_0 \bra\cp_{j_2}\cp_1 \ket_0\eta^{i_1j_1}\eta^{i_2j_2}+\cr&
\bra\overbrace{\cp_2\cp_{i_1}\cp_{i_2}} \ket_0
\bra\cp_{j_1} \cp_1  \ket_0 \bra\cp_{j_2} \cp_2 \ket_0\eta^{i_1j_1}
\eta^{i_2j_2}+
\bra\overbrace{\cp_2\cp_{i_1}\cp_{i_2}}   \ket_0
\bra\cp_{j_1}   \ket_0 \bra\cp_{j_2}  \cp_2\cp_1 \ket_0\eta^{i_1j_1}
\eta^{i_2j_2}+\cr&
\bra\overbrace{\cp_2\cp_{i_1}\cp_{i_2}}  \ket_0
\bra\cp_{j_1} \cp_1 \cp_2 \ket_0 \bra\cp_{j_2}  \ket_0\eta^{i_1j_1}
\eta^{i_2j_2},\cr}
\eq{ex221}
$$
where an implicit sum over $i_1$, $j_1$, $i_2$ and $j_2$ is understood.
Basically, we have a situation similar to Figure 4. All we do is distribute
the non-marked operators in the correlation function in all possible ways.
Note the two inverse metrics is each term: those are from the double 
insertion of the identity operator.

To continue we need to evaluate the contact terms. Using ghost number
conservation, we translate the contact term into one operator of the same
ghost number, with some yet to be determined coefficient. Thus, for example,
$$\overbrace{\cp_2\cp_a\cp_b} = \beta_{2 a b}\cp_{2+a+b-2(q+1)}. 
\eq{conprop}$$
 Assume
that $\beta_{2 a b}$ is proportional to $|a||b|$. This assumption, that the
contact term is proportional to the dimensions of the other (non-marked)
operators, is not surprising if we remember that the two-point function is
proportional to the dimension. In this way the contact coefficient will
cancel the contributions from the inverse metric $\eta^{i_1j_1}$. 

Now, looking at equation \putlab{ex221}, we see that the proportionality 
constant in the contact terms, if not zero, will cancel out of the 
equation. Of course, if it is
zero, then the theory is trivial. Thus, the exact numerical value of the 
contact coefficient is irrelevant. Only the fact that it's proportional 
to the dimensions of the non-contact operators is important.

One can easisly convice oneself the the cancellation of the numerical
coefficient in the contact term happens for any $(1,q)$ model. We thus
will define the constant to be $1$ and we get the following contact algebra,
$$\overbrace{\cp_n\cp_{i_1}\ldots\cp_{i_m}} = |i_1|\cdots|i_m|
 \cp_{n+\sum_{j=1}^m i_j - m(q+1)}. \eq{contactterms}$$

Returning to our example,
using the ghost number conservation rule, the contact term value and the 
metric, we can sum over the indices in equation \putlab{ex221}. We get,
$$\eqalign{ 0 = &
\bra \cp_2\cp_2\cp_1  \ket_0 \bra \cp_{-5}\ket_0 \bra \cp_{-5}\ket_0+
\bra \cp_{-2}\cp_2  \ket_0 \bra \cp_{-1}\cp_1\ket_0 \bra \cp_{-5}\ket_0+\cr &
\bra \cp_{-2}\cp_2  \ket_0 \bra \cp_{-5}\ket_0 \bra \cp_1\cp_{-1}\ket_0+
\bra \cp_{-1}\cp_1  \ket_0 \bra \cp_2\cp_{-2}\ket_0 \bra \cp_{-5}\ket_0+\cr &
\bra \cp_{-1}\cp_1  \ket_0 \bra \cp_{-5}\ket_0 \bra \cp_2\cp_{-2}\ket_0+
\bra \cp_{-5}  \ket_0 \bra \cp_2\cp_{-2}\ket_0 \bra \cp_1\cp_{-1}\ket_0+\cr &
\bra \cp_{-5}  \ket_0 \bra \cp_1\cp_{-1}\ket_0 \bra \cp_2\cp_{-2}\ket_0+
\bra \cp_{-5}  \ket_0 \bra \cp_{-5}\ket_0 \bra \cp_2\cp_2\cp_1\ket_0+\cr &
\bra \cp_{-5}  \ket_0 \bra \cp_2\cp_2\cp_1\ket_0 \bra \cp_{-5}\ket_0,\cr }
\eq{ex221b}$$
and substituting the values $\bra\cp_1\cp_{-1}\ket_0=\bra\cp_{-1}\cp_{1}\ket_0
=1$, $\bra\cp_2\cp_{-2}\ket_0=\bra\cp_{-2}\cp_{2}\ket_0=2$ and $\bra\cp_{-5}
\ket_0 = -4$, one gets the answer,
$$\bra\cp_2\cp_2\cp_1\ket_0 = 1. \eq{ex221ans}$$

It is instructive to compute the same correlation function when $\cp_1$ is
the marked operator. In this case there is only one degeneration ($\cp_1$ is
the puncture operator!), and the degeneration equation is
$$\eqalign{ 0 = \sumdel \bra\cp_1 & \cp_2\cp_2\ket_0 = \cr&
\sum_{i_1,j_1}\bra\overbrace{\cp_1\cp_{i_1}}  \cp_2\cp_2 \ket_0
\bra\cp_{j_1}  \ket_0\eta^{i_1j_1}+\cr&
 2 \sum_{i_1,j_1}\bra\overbrace{\cp_1\cp_{i_1}}  \cp_2 \ket_0
\bra\cp_{j_1}  \cp_2\ket_0\eta^{i_1j_1}+\cr &
\sum_{i_1,j_1}\bra\overbrace{\cp_1\cp_{i_1}}  \ket_0
\bra\cp_{j_1} \cp_2 \cp_2\ket_0\eta^{i_1j_1}\cr}. \eq{ex122}$$
Notice the combinatorial factor in the second term. Continuing in a similar
fashion to equation (\putlab{ex221b}), one get the same result.

It is clear that the anti-states are very important here. The fact
that the one-point correlation $\bra\cp_{-q-1}\ket_0$ is non-vanishing
guarantees that by performing the sum over degenerations on a correlation
function, the same correlation function will reappear in the sum.
There will always be terms in the sum in which all the 
other (\ie, non-marked) operators will be on one of the surfaces (spheres).
By ghost number conservation, the last operator on this surface that either
came from the contact term or from the identity operator is guaranteed
to be identical to the original marked operator. In equation (\putlab{ex221b}),
for example, the first, eighth and ninth terms are of this kind.

\subsection{3. The General Skeletal Degeneration Equation}

As discussed above, one notes that the marked operator is the only element
that specifies the number of degenerations. Also, one notes, that that
operator comes in contact {\it only} with the degenerations, which are
represented by the identity operator. Thus we can write a skeletal 
degeneration equation that depends only on the first operator. We will 
write, for the general $(1,q)$ model with $q-1$ primary fields,
$$\sumdel \bra\sigma_n(\cp_\alpha) X\ket_0 =\sum_{\bigcup_{i=0}^\alpha 
X_i = X}
\bra\ldb\sigma_n(\cp_\alpha)\cp_{-i_1}\ldots\cp_{-i_\alpha}\rdb X_0 \ket_0
\bra \cp_{i_1}X_1\ket_0\cdots\bra\cp_{i_\alpha} X_\alpha\ket_0=0, \eq{general}$$
where we defined the normalized contact 
$$\ldb\cp_n\cp_{i_1}\ldots\cp_{i_m}\rdb = \cp_{n+\sum_{j=1}^m i_j - m(q+1)}.
\eq{normcont}$$
Note that we didn't write the inverse metric factors in equation 
(\putlab{general}). Those cancel with the
regular contact term and yield the normalized contact term. That is
$$ \overbrace{\cp_n\cp_{i_1}\ldots\cp_{i_m}} = |i_1|\cdots|i_m|\ldb
 \cp_n\cp_{i_1}\ldots\cp_{i_m}\rdb. \eq{reg2norm}$$
 There is an
implicit summation in equation \putlab{general} over the $i_j$'s. However, because of
ghost number conservation, there will be only one possible $i_j$. The `$X$'
in equation (\putlab{general}) is some arbitrary collection of operators 
and the various $X_j$ have no element in common.

\section{Recursion Relations and Effective Contacts}

In equation (\putlab{general}), there are $(\alpha +1)$ terms in the sum
that are proportional to the original correlation function, as explained 
in the end
of subsection 3.2. Computing these terms separately we can transform
the degeneration equation into a recursion relation. We will get
the following skeletal recursion relation
$$ \bra\sigma_n(\cp_\alpha) X\ket_0 =-{{1}\over{(\alpha+1)(-q)^\alpha}}
\sum_{\bigcup X_i = X \atop X_i \ne X}
\bra\ldb\sigma_n(\cp_\alpha)\cp_{-i_1}\ldots\cp_{-i_\alpha}\rdb X_0 \ket_0
\bra \cp_{i_1}X_1\ket_0\cdots\bra\cp_{i_\alpha} X_\alpha\ket_0. \eq{rec}$$

For $\alpha=1$, for example, the above equation is
$$ \bra\sigma_n(\cp_1) X\ket_0 ={{1}\over{2q}}
\sum_{X_0\bigcup X_1 = X \atop X_i \ne X}
\bra\ldb\sigma_n(\cp_1)\cp_{-i_1}\rdb X_0 \ket_0
\bra \cp_{i_1}X_1\ket_0, \eq{rec1}$$
which is identical to the Virasoro constraints restricted to the sphere.

It is interesting to note that in equation \putlab{rec1} we recover the
contact algebra as written by the Verlindes. (Actually, this is slightly
more general, since the collection of operators, $X$, can contain operators
that are not descendants of the puncture operator, $\cp_1$.) Explicitly
writing $X$ and evaluating the possible two-point functions, we get
$$\eqalign{\bra\sigma_n(\cp_1)\cp_{i_1}\ldots\cp_{i_m}\ket_0 = &{{1}\over{q}}
\sum_{j=1}^m i_j \bra \sigma_{n-1}(\cp_{i_j})\cp_{i_1}\ldots
\cp_{i_{j-1}} \cp_{i_{j+1}}\ldots\cp_{i_m}\ket_0 +\cr &{{1}\over{2q}}\sum_{X_0\bigcup X_1 = X\atop X_i\ne\emptyset,X_i\ne\{\cp_l\}
\forall j} \bra\ldb\sigma_n(\cp_1)\cp_{-i_1}\rdb X_0 \ket_0\bra \cp_{i_1}X_1\ket_0.\cr}\eq{rec2}$$
The first sum on the right hand side is a result of splitting $X$ into a 
single operator and the rest, and since this single operator can be in any 
of the two surfaces, we get the factor of two, and hence the value $1/q$ in
front. Also note the factor of $i_j$ in the first term. That is a result
of evaluating the two-point function.
 The second term on the right hand side is the rest of the possible
divisions of the set $X$. In the language of the Verlindes we identify
the first sum as the contribution from the contact terms
and the second sum as degenerations of the surface.
We will refer to these (that is-Verlindes) contact terms as {\it effective}.

Similarly, for $\alpha=2$ we can get an effective contact algebra for
the second primary field. It is interesting to note that although in the
degeneration equation (\putlab{general}) the operator $\cp_2$ has contacts
with $2$ other operators (no more and no less), the effective contact algebra
has single {\it and} double contacts.

Let us demonstrate this by calculating a less trivial example.
Consider the $(1,3)$ model which has two primary fields, $\cp_1$ and
$\cp_2$, with descendants, $\cp_{3i+1}$ and $\cp_{3i+2}$, 
respectively. 
Now using the above prescription, we can write a degeneration equation to
compute a correlation function involving $\cp_{3i+2}$ and $\cp_1$'s.
$$\sum_{a=0}^{i+2}\sum_{b=0}^{i+2-a}{i+2\choose a}
 {i+2-a\choose b}\bra \ldb\cp_{3i+2}\cp_{-i_1}\cp_{-i_2}\rdb
\cp_1^a\ket_0\bra \cp_{i_1} \cp_1^b\ket_0
\bra \cp_{i_2}\cp_1^{i+2-a-b}\ket_0 = 0 . \eq{ex1}$$
Separating the sum to the cases for which $a$, $b$ and/or $i+2-a-b$ are zero
or one, we get a recursion relation without anti-states,
$$\eqalign{\bra &\cp_{3i+2}  \cp_1^{i+2}\ket_0={{2}\over{3}}(i+2)
\bra \cp_{3(i-1)+2}  \cp_1^{i+1}\ket_0 -{{1}\over{9}}(i+2)(i+1)
\bra \cp_{3(i-2)+2}  \cp_1^i\ket_0  \cr &+
{{2}\over{9}}\sum_{a=2}^{i}{i+2\choose a}
\bra \cp_{3(a-2)+2}\cp_1^a\ket_0\bra \cp_{3(i-a)+2}\cp_1^{i+2-a}\ket_0 \cr&-
{{2}\over{27}}(i+2)\sum_{a=2}^{i+1}{i+1\choose a}
\bra \cp_{3(a-2)+2}\cp_1^a\ket_0\bra \cp_{3(i-a-1)+2}\cp_1^{i+1-a}\ket_0 \cr &-
 {{1}\over{27}}\sum_{a=2}^{i}\sum_{b=2}^{i-a}{i+2\choose a}{i+2-a\choose b}
\bra  \cp_{3(a-2)+2}\cp_1^a\ket_0\bra \cp_{3(b-2)+2}\cp_1^b\ket_0
\bra \cp_{3(i-a-b)+2}\cp_1^{i+2-a-b}\ket_0 .\cr} \eq{ex1cont}$$
The terms in this equation are interpreted in the following way. The first
is an effective contact between the marked operator and (each) one
of the other operators. The second term is a double contact with two of them.
The third term is a single contact of the marked operator with a single
surface degeneration and the forth is a double contact with one operator
{\it and} one degeneration. The last term is a contact with double 
degeneration.
 
One can extract the effective contact between $\cp_{3i+2}$
and $\cp_1$. It is
$$\lll\cp_{3i+2}\cp_1\rrr={2\over 3}\cp_{3(i-1)+2},\eq{effcont}$$
where we introduced a notation for the effective contact.
Similarly, the effective double contact between $\cp_{3i+2}$ and 
two $\cp_1$'s is
$$\lll\cp_{3i+2}\cp_1\cp_1\rrr=-{2\over 9}\cp_{3(i-2)+2}.\eq{effcont2}$$

It is instructive to compare equations \putlab{ex1} and \putlab{ex1cont}.
The second appears to be far more complicated, although it can be derived
from the first one. The $W_3$ constraints will give rise to such an equation.
(See Appendix E.)
 
Following the same procedure, we can compute more general 
recursion relations. We cannot compare them
with the results from $W_q$ constraints ($q>3$), since no explicit
results are available. 
 
A generalization of equation (\putlab{ex1cont}) to the $(1,q)$ model 
is given by the following,
$$\eqalign{\bra&\cp_{qi+q-1}\cp_1^{i+2}\ket_0=\alpha(2)\sum_{a=0}^{i+2}
{i+2\choose a}\bra\cp_{qa-q-1}\cp_1^a\ket_0\bra\cp_{q(i+2-a)-q-1}\cp_1^{i+2-a}
\ket_0 \cr
&+\alpha(3)\sum_{a=0}^{i+2}\sum_{b=0}^{i+2-a}{i+2\choose a}{i+2-a\choose b}
\bra\cp_{qa-q-1}\cp_1^a\ket_0\bra\cp_{qb-q-1}\cp_1^b\ket_0
\bra\cp_{q(i+2-a-b)-q-1}\cp_1^{i+2-a-b}\ket_0 \cr
&+\ldots\cr
&+\alpha(k)\sum_{a_1=0}^{i+2}\sum_{a_2=0}^{i+2-a_1}\cdots\sum_{a_{k-1}=0}^{i+2-
\sum_{j=1}^{k-2} a_j}
{i+2\choose a_1}{i+2-a_1\choose a_2}\cdots{i+2-\sum_{j=1}^{k-2}
a_j\choose a_{k-1}} \times\cr
&\,\,\bra\cp_{qa_1-q-1}\cp_1^{a_1}\ket_0\bra\cp_{qa_2-q-1}
\cp_1^{a_2}\ket_0\cdots
\bra\cp_{qa_{k-1}-q-1}\cp_1^{a_{k-1}}\ket_0\bra\cp_{q(i+2-\sum_{j=1}^{k-2}
a_j)-q-1}\cp_1^{i+2-\sum_{j=1}^{k-1}a_j}\ket_0 \cr
&+\ldots\cr} \eq{ex2}$$
where the coefficients $\alpha(k)$ are
$$\alpha(k)={(-1)^k\over k!}\prod_{j=1}^{k-1}(1-{j\over q}).\eq{alphas}$$

Another formula we will need in order to derive the effective contacts
is one involving three distinct primary fields. This can be similarly
derived,
$$\eqalign{
&\bra\cp_{qi+\alpha}\cp_{qj+q-\alpha}\cp_1^{i+j+1}\ket_0=
\cr &+2\beta(2)
\sum_{a=0}^{i+j+1}{i+j+1\choose a}\bra\cp_{aq-qi-q-\alpha}\cp_{qj+q-\alpha}
\cp_1^a\ket_0\bra\cp_{q(i+j+1-a)-q-1}\cp_1^{i+j+1-a}\ket_0 \cr
&+3\beta(3)\sum_{a_1=0}^{i+j+1}\sum_{a_2=0}^{i+j+1-a_1}{i+j+1\choose a_1}
{i+j+1-a_1\choose a_2}\times\cr
&\bra\cp_{a_1q-qj-q+\alpha}\cp_{qj+q-\alpha}\cp_1^{a_1}
\ket_0\bra\cp_{qa_2-q-1}\cp_1^{a_2}\ket_0\bra\cp_{q(i+j+1-a_1-a_2)-q-1}
\cp_1^{i+j+1- a_1-a_2}\ket_0 \cr &+ \ldots\cr
&+k\beta(k)\sum_{a_1=0}^{i+j+1}\cdots\sum_{a_{k-1}=0}^{i+j+1-\sum_{j=1}^{k-2}
a_j}{i+j+1\choose a_1}\cdots{i+j+1-\sum_{j=1}^{k-2}a_j\choose a_{k-1}}\times
\cr
&\bra\cp_{a_1q-q j-q+\alpha}\cp_{qj+q-\alpha}\cp_1^{a_1}\ket_0
\bra\cp_{qa_2-q-1}\cp_1^{a_2}\ket_0\cdots\bra\cp_{qa_{k-1}-q-1}
\cp_1^{a_{k-1}}\ket_0 \times \cr
&\bra\cp_{q(i+j+1-\sum_{j=1}^{k-1}a_j)-q-1}
\cp_1^{i+j+1-\sum_{j=1}^{k-1}a_j}\ket_0 +\ldots\,.\cr}\eq{ex3}$$
where
$$	\beta(k)={(-1)^k\over k!q^{k-1}}\prod_{l=0}^{k-2}(\alpha-l). 
\eq{betas}$$

We evaluate the effective
contact terms derived from equations (\putlab{ex2}) and (\putlab{ex3}).
First, using equation (\putlab{ex2}) we can show that,
$$\eqalign{\lll\cp_{qi+q-1}\cp_1\rrr&={q-1\over q}\cp_{qi-1}, \cr
       \lll\cp_{qi+q-1}\cp_1\cp_1\rrr&=-{(q-1)(q-2)\over q^2}\cp_{qi-q-1},\cr
                            &\,\,\vdots \cr
 \lll\cp_{qi+q-1}\underbrace{\cp_1\cdots\cp_1}_{k-1\ {\rm times}}\rrr&=
(-1)^k\prod_{j=1}^{k-1}(1-{j\over q})\cp_{q(i-k+2)-1}. \cr}\eq{ex2cont}$$
From equation (\putlab{ex3}) we get the following effective contact
terms,
$$\eqalign{\lll\cp_{qi+\alpha}\cp_{qj+q-\alpha}\rrr&={\alpha\over q}(qj+q-\alpha)
\cp_{qi+qj-1}, \cr
\lll\cp_{qi+\alpha}\cp_1\rrr&={\alpha\over q}\cp_{qi+\alpha-q}, \cr
\lll\cp_{qi+\alpha}\cp_{qj+q-\alpha}\cp_1\rrr&=-{\alpha(\alpha-1)\over
 q^2}(qj+q-\alpha) \cp_{q(i+j)-q-1},\cr }\eq{ex3cont}$$
and so forth.

By looking at more complicated recursion relations with more distinct 
primary fields, we conjecture that in general the effective
contact algebra is given as follows,
$$\eqalign{\lll\cp_i\cp_j\rrr&=1!{\alpha\choose 1}{j\over q}\cp_{i+j-(q+1)},\cr
\lll\cp_i\cp_j\cp_k\rrr&= -2!{\alpha\choose 2}{jk\over
 q^2}\cp_{i+j+k-2(q+1)},\cr &\vdots \cr
\lll\cp_i\prod_{k=1}^n\cp_{j_k}\rrr&= (-1)^{n-1} n!{\alpha\choose n}
{\prod_{k=1}^nj_k\over q^n}\cp_{i+\sum_{k=1}^nj_k-n(q+1)},\cr}\eq{contacts}$$
where $\alpha=i\mod q (>0)$.

One may explore what kind of algebra one gets by considering
the commutators of the two-term contacts in (\putlab{contacts}).
We get the following result,
$$\lb \cp_i,\cp_j\rb={1\over q}\left(\left(i\mod q\right) j-
\left(j\mod q\right)i\right)
\cp_{i+j-q-1}.\eq{commut}$$
If $q=2$, we simply get
$$\lb \cp_i,\cp_j\rb={1\over 2}(j-i)\cp_{i+j-3}.$$
For $q=3$ let us define $\hat\cq_i \equiv \cp_{3i+2}$ and $\hat\cp_j
\equiv \cp_{3j+1}$. Then,
$$\eqalign{\lb \hat\cp_i,\hat\cq_j\rb&={1\over 3}(2j-i)\hat\cq_{i+j-4},\cr
           \lb \hat\cp_i,\hat\cp_j\rb&={2\over 3}(j-i)\hat\cp_{i+j-4},\cr
           \lb \hat\cq_i,\hat\cq_j\rb&=0,\cr}\eq{walg}$$
The last commutator follows from the fact that there is no operator
with the correct ghost number.

\section{The Torus and Higher Genus}

It is natural to investigate the generalization to higher genus. It
turns out that there is a natural extension of the degeneration equation
in higher genus. One problem, however, does arise because of a
counting ambiguity
in ordering the operators in the identity insertion. Nevertheless,
we will be able to derive partial results, whenever this ambiguity does
not occur. Our discussion will be mostly restricted to genus $1$, but
we will also make some comments on higher genus at the end.

\subsection{1. Notation and Diagrams}

We begin by naturally extending the various diagrams of Section 3.1 to
tori. A correlation function on a genus one surface, $\bra\cp_n\cp_{i_1}\ldots
\cp_{i_m}\ket_1$, is depicted in \Figure{gen1}. Again we use an $\times$
symbol for the marked operator, $\cp_n$. 

\putfig{gen1}{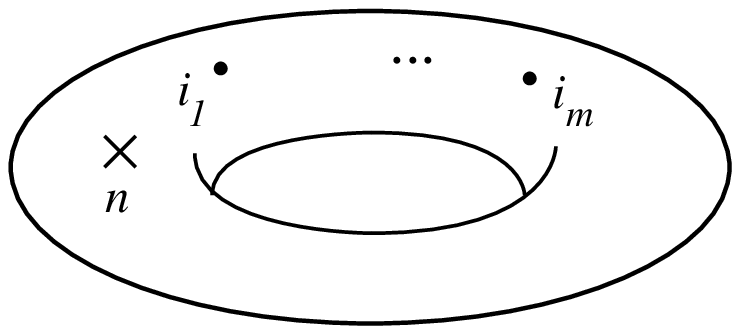}{The Correlation Function $\bra\cp_n\cp_{i_1}
\ldots\cp_{i_m}\ket_1$.}

Contacts between operators will be represented by lines, as before (see
Fig.~2). The difference in higher genus arises in the degenerations. 
In genus $1$ there are two possible
types of degenerations. The first is the splitting of a sphere 
off the torus, and the
second is a handle being pinched off. One should notice, however, that
the marked operator can end up in the sphere being split off (or, of course,
remain in the torus). \Figure{split} is the degeneration of the first
kind, a sphere being split off. The correlation function $\bra\cp_1 X \ket_1$
has a sphere split off and the marked operator can end up in two surfaces.

\putfig{split}{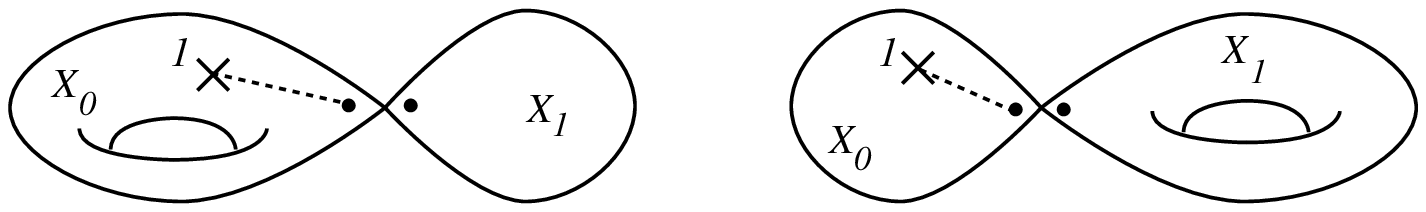}{Two Possibilities for the Marked Operator after 
a Split.}

A handle pinch in the same correlation function is shown
is \Figure{pinch}. Notice the insertion of the identity operator.

\putfig{pinch}{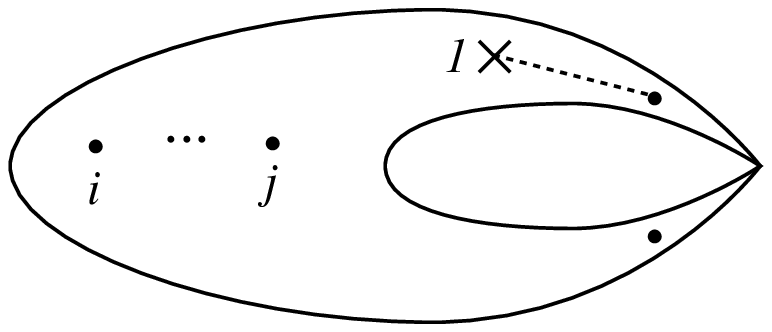}{A Handle Pinch in $\bra\cp_1 \cp_i\ldots\cp_j
\ket_1$.}

One notes, that when a handle is pinched, there two operator from the
identity insertion end up in the same surface (in our case, a sphere).
This actually increases the number of operators in the correlaiton function,
but decreases the genus.

\subsection{2. The Degeneration Equation at Genus One}

Formally, the degeneration equation is exactly the same as before; \ie,
the sum over all degenerations in contact with the marked operator is zero.
However, one has to remember that there is also the possibility of a handle
pinch. The easiest way to visualize the degeneration equation is to 
draw a skeletal diagram, in which the marked operator will be in contact
with the appropriate identity operator, and the sum over different
{\it types} of degenerations will be explicitly present. All there is
to do then is to distribute the other (non-marked) operators and
calculate.

Let us demonstrate it for the simplest case for which the marked operator
is a descendant of $\cp_1$. Such an operator, as explained above, can
have only one contact. Thus, the skeletal degeneration equation will be,
$$\sumdel \bra\cp_{qn+1} * \ket_1 = \bra\ldb\cp_{qn+1}\cp_{-i_1}\rdb *\ket_1
\bra\cp_{i_1} *\ket_0+\bra\ldb\cp_{qn+1}\cp_{-i_1}\rdb *\ket_0
\bra\cp_{i_1} *\ket_1 + \bra\ldb\cp_{qn+1}\cp_{-i_1}\rdb\cp_{i_1} *\ket_0=0.
\eq{gen1skel}$$
In this equation, the ``$*$'' in the left hand side is any collection of 
physical operators and the ``$*$'' on the right hand side 
is an implicit summation over all the possible distributions of
these operators into the various surfaces. Diagrammatically, the first
two parts are depicted in Fig.~6 and the third part in 
Fig.~7.

In general, for descendants of higher primary fields, there will be more
then one contact (and degeneration). In that case the specified number of
degenerations could be a combination of a pinch and multiple splittings
or only splittings\footnote[*]{It might be obvious here how to generalize
this to higher genus.}.

\subsection{3. An Example Computation}

As an example, let us compute the correlation function
$\bra\cp_{q(i+1)+1}\cp_1^i\ket_1$ when $i > 0$.
Using the degeneration equation (\putlab{gen1skel}), we get,
$$\eqalign{\sumdel \bra\cp_{q(i+1)+1}& \cp_1^i \ket_1 = \sum_{a=0}^{i}
{i\choose a}\bra\ldb\cp_{q(i+1)+1}\cp_{-j}\rdb \cp_1^a\ket_1
\bra\cp_{j} \cp_1^{i-a}\ket_0+ \cr &\sum_{a=0}^{i}
{i\choose a} \bra\ldb\cp_{q(i+1)+1}\cp_{-j}\rdb \cp_1^a\ket_0
\bra\cp_{j} \cp_1^{i-a}\ket_1+ \bra\ldb\cp_{q(i+1)+1}\cp_{-j}\rdb\cp_{j} \cp_1^i\ket_0=0.\cr}
\eq{part1}$$
As before, there is an implicit summation over the index $j$ (from the 
identity operator). In the first two terms, as before, there is only one
value of $j$ for which the correlation function will not vanish (from ghost
number conservation). In the third term, however, there is a range of
possible $j$'s for which the correlation will not vanish. This is exactly
the point where we have the ambiguity problem with normal ordering.
It is true that as long as the range of possible $j$'s is finite,
the computation can proceed as usual and the result is totally consistent
with the KdV approach. It is when the range of $j$ is infinite that a 
problem arises. At first glance
the counting ambiguity seems never to occur, but, alas,
this is not true. In the above example it occurs for $i=0$ (only!). In fact,
one can easily convince oneself that 
only when the marked operator is a descendant 
of $\cp_1$ do we have an ambiguity, since this is the only one point
correlation function. For higher primary fields, however, there can be
a handle pinch together with sphere splitting. In that case the infinity
problem will arise again. Our method works whenever there is no such
ambiguity. It breaks down otherwise.

Continuing with the above example (and limiting ourselves to $i>0$), 
we can evaluate the contact terms and compute the possible $j$'s. We
get,
$$\eqalign{ \sum_{a=0}^{i}
{i\choose a}\bra\cp_{q(a+1)+1} \cp_1^a\ket_1
\bra\cp_{q(i-a-2)+q-1}& \cp_1^{i-a}\ket_0+ \cr \sum_{a=0}^{i}
{i\choose a}\bra\cp_{q(a-2)+q-1} \cp_1^a\ket_0
\bra\cp_{q(i-a+1)+1} \cp_1^{i-a}\ket_1 &+ \sum_{j=0}^{q i}
\bra\cp_{qi-j}\cp_{j} \cp_1^i\ket_0=0.\cr}
\eq{part2}$$
Transforming this equation into a recursion relation, as was done on the
sphere, we can compute the result explicitly. It agrees with the KdV result
which is
$$\bra\cp_{q(i+1)+1}\cp_1^i\ket_1 = {{q-1}\over{24}} 
\prod_{j=1}^{i+1}(j + {{1}\over{q}}). \eq{part3}$$

Note, though, that in equation (\putlab{part3}) one {\it can} set $i=0$.
The result is
$$\bra\cp_{q+1}\ket_1 = {{q^2-1}\over{24 q^2}}. \eq{dilaton22}$$ 
It leads one tho think that there might be a way to regularize the infinite
summation in the degeneration equation and get a meaningful result. We
have been able to get some partial result, but we haven't been able to
solve the general problem.

\subsection{4. Higher Primary Fields on the Torus}

When the marked operator is a descendant of a higher primary field,
there are two points that one has to be careful with. The first
is combinatorics. One has to count carefully the number of possible
degenerations. The second point to notice is the distribution of
the operators from the identity operator over the different surfaces
when there is a pinch and a split. There are two 
possibilities to perform a pinch and a degeneration on a torus. The
first is to first pinch the handle and then to pinch a sphere from
the side. The second is to pinch the handle in two places at once.
In \Figure{twoway} those two possibilities are depicted for the
correlation function $\bra\cp_2\cp_k\ldots\cp_l\ket_1$.

\putfig{twoway}{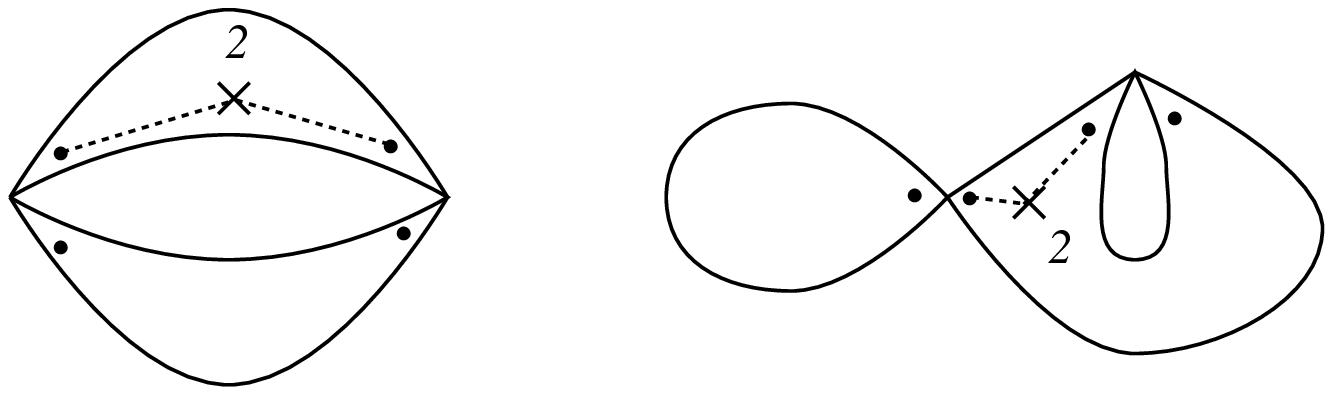}{Two Ways for a Pinch and a Split.}

It is interesting to note that for the $(1,3)$ model, we can compute
many correlation functions at genus one, as we do not get infinite summations.

\subsection{5. Higher Genus}

In principle, the degeneration equation can be generalized to higher genera
(larger than $1$). However, the regularization ambiguity limits the
results, as the probability of encountering it grows at higher genus. 

It is true, nevertheless, that if the marked operator is a descendant
of the puncture operator $\cp_1$, then the degeneration equation
exists. Of course, after the first recursion, one might find that one
needs to compute correlation functions with other marked operators
that may have the ambiguity problem. For the $(1,2)$ model which has only
one primary field any correlation can be computed.
For the $(1,3)$ model, we also suspect that any correlation function can 
be computed, but this is due to a coincidence that the infinite summation
does not occur. An example of a correlation function with normal ordering
problem is $\bra\cp_3\cp_7\ket_1$ in the $(1,4)$ model.

\section{Conclusions}

In this paper we used some of the general results about the correlation
functions of the $(1,q)$ models
derived from the KdV hierarchy\ref{mont} in order to achieve a better 
understanding of the structure of quantum gravity. A simple underlying
structure for the correlation functions of the $(1,q)$ models on the
sphere was discovered. It is applicable to models with an arbitrary finite
number of primary fields. We introduced an ``identity'' operator (which in
quantum theories of gravity is a difficult concept to define) and
computed correlation functions by inserting it at the surface degenerations.
This is intuitively how one would have expected to compute correlation
functions in a topological quantum theory of gravity. We demonstrated 
the immense
simplicity of this procedure over that based on the $W_n$ constraints.
Any correlation function on the sphere could be computed directly 
including those of primary fields. 

The introduction of a formal adjoint operation and a local identity
operator may hint at possible extensions to models of gravity with
a continuous number of ``primary fields.''  The operators of negative
dimension which were introduced in Section 2  
to simplify computations may turn out to have
some physical interpretation. They certainly cannot be interpreted
as forms on the relevant moduli space (that of Riemann surfaces for
$q=2$).  

Since we provide explicit expressions for correlation functions as
functions of $q$, 
an interesting open question which may be considered using the 
results of this paper and References [\putref{mont}] and [\putref{rivl2}] 
is to consider the limit $q\rightarrow\infty$. This should correspond
to some semi-classical limit with $c=-\infty$, depending on the way 
the limit is taken.
It is clear that if the limit is taken
appropriately some of the correlation functions will converge
to a finite value. It is not transparent, however, how the fields
$\cp_i$ relate to the Liouville field if $i$ becomes a
continuous parameter as $q\rightarrow\infty$. 

Another problem is the counting ambiguity at genus one
(and higher). A few possibilities to cure this problem are
currently under investigation. One is to find an appropriate regularization
scheme. Another 
is to investigate the possibility that the anti-states do not commute
with the regular physical states. The analogy might be with an infinite
set of harmonic oscillators. For each oscillator we have a creation and
annihilation operators denoted by $a_i^\dagger$ and $a_i$ that do not
commute. The physical operators $\cp_i$ can be thought of as creation
operators and the anti-states as annihilation ones. All the creation 
operators mutually commute as do the annihilation operators. There is, however,
a non-zero commutator between a creation operator and its conjugate.
A third possibility is to extend the definition of the identity
operator and hence the metric. We used only the two-point function
on the sphere to construct the identity. It is possible that two-point
functions on higher genus will contribute to the identity. This
should not change the genus $0$ results, but may change higher genus
answers. All these options are currently under investigation.

\bigskip
\lefttext{\bf Acknowledgments}
\smallskip
G.~R.\ would
especially like to thank Alice Quillen for numerous
discussions and for helping to write some crucial computer programs.
D.~M.\ would especially like to thank Cobi Sonnenschein for very
fruitful discussions and Tel Aviv University for its hospitality
where some of this work was done. We would also both like to
thank John Schwarz for reading
the manuscript.

The computations for this paper where done using Mathematica and
C programs on a NeXTstation. The co-author of the programs (Alice
Quillen) has agreed to make them available to interested readers 
upon request. They include a Mathematica program that performs 
operations on general pseudo-differential operators, 
another Mathematica program that computes pseudo-differential
operators in the $(1,q)$ models and a C program with a Mathematica
interface (using MathLink) that computes correlation functions
according to the method presented in this paper.

D.~M.'s work was supported in part by the U.S.
Department of Energy under Contract No. DE-AC0381-ER40050. G.~R.'s
work was supported in part by the
Director, Office of Energy Research, Office of High Energy and Nuclear
Physics, Division of High Energy Physics of the U.S. Department
of Energy under contract DE-AC03-76SF00098 and in part by the
National Science Foundation under grant PHY90-21139.

\sectionnum=0 
\sectionnumstyle{Alphabetic} 

\appendix{KdV Gravity}

The KdV approach to two-dimensional quantum gravity is a recipe for
computing correlation functions using the generalized KdV hierarchy
of differential equations. It is based on a $q$th order differential
operator,
$$Q= D^q + u_{q-2}(x,{\bf t}) D^{q-2} +  u_{q-3}(x,{\bf t}) D^{q-3}
+ \cdots + u_0(x,{\bf t}), \eq{Qop}$$
where $u_i$ are formal power series in $x$ and an infinite number of
`time' variables, ${\bf t} = (t_1,t_2,\ldots)$. $D$ is the differentiation
operator $\partial/{\partial x}$. The dependence of these
functions on the time variables are determined by
the KdV hierarchy (the KdV `flows')
$${{\partial Q}\over{\partial t_i}}=\left[ R^i_+, Q\right],
\eq{kdvhierarchy}$$
where $R$ is a the pseudo-differential operator satisfying $R^q=Q$, and
$R_+$ is the differential part of $R$.
The dependence on $x$ is specified by the so-called `string equation',
$$\left[ P, Q \right ] = 1, \eq{stringequation}$$
where $P$ is a $p$th order differential operator. One can show that
$P$ must be a linear combination of various $R^i_+$. If $P=R^q_+$,
then we have the $(p,q)$ model.

To compute the correlation functions one maps the fields in the theory
to powers of $R$,
$$\cp_i \leftrightarrow R^i, \eq{map}$$
and identifies the two puncture correlation function to be proportional
to $u_{q-2}$. That is,
$$\bra \cp_1\cp_1 \ket = \Res_{-1} R, \eq{p1p1}$$
where $\Res_i\co$ is the coefficient of $D^i$ in the operator $\co$.
One can imagine that the original
Lagrangian is perturbed by $\sum_i t_i\cp_i$ where the sum is taken over
all possible fields.
Taking a partial derivative with respect to $t_i$
will insert the operator $\cp_i$ into the correlation function. The basic idea
is then to identify these parameters with the flows of the KdV hierarchy!
Thus one can compute,
$$\bra\cp_1\cp_1\cp_m\ket
={{\partial}\over{\partial t_m}}\bra\cp_1\cp_1\ket= \Res_{-1}\lb R^m_+,R \rb.
\eq{p1p1pm}$$
After some manipulation one gets
$$\bra\cp_1\cp_1\cp_m\ket=\left(\Res_{-1}R^m\right)', \eq{p1p1pmstar}$$
and integrating
$$\bra\cp_1\cp_m\ket = \Res_{-1} R^m + \alpha_m. \eq{p1pm}$$
This integration constant is usually set to zero. 
However, when discussing the $(1,q)$ models, we will
return to this point.

Proceeding, one can compute
$$\bra\cp_1\cp_m\cp_n\ket = \Res_{-1}\lb R^m_+,R^n \rb, \eq{p1pmpn}$$
and similarly
$$\bra\cp_1\cp_l\cp_m\cp_n\ket =
\Res_{-1}\left(\lb \lb R^l_+, R^m \rb_+,R^n\rb+\lb R^m_+,\lb R^l_+,
R^n\rb\rb\right).  \eq{p1plpmpn}$$

In a generic model there are some features which are independent
of the exact choice of the string equation (\putlab{stringequation}).
First, as one can one easily see, inserting the operator $\cp_{m q}$
into the correlation function is trivial,\ie, the correlation function
vanishes because $R^{m q}_+ = Q^m_+ \equiv Q^m $ and thus
commutes with any power of $R$. Thus, naturally, the operators
come in bands of $q-1$.They are identified with primary
and descendant fields in the following way: The primaries are $P_\alpha$
for $1 \le \alpha \le q-1$ and the $m$th descendant of $\cp_\alpha$
is $\sigma_m(\cp_\alpha)=\cp_{m q +\alpha}$.

Second, the fact the the KdV hierarchy is completely integrable
guarantees that the correlation functions are independent of the
order of the operators, since the flows (\putlab{kdvhierarchy})
commute.

Third, the correlation functions are non-perturbative. That is, they
include the sum over all genera.

Fourth, usually one cannot write the solution in a closed analytic
form. This is because, generically, the string equation is horribly
non-linear.

However, there is a set of models, the $(1,q)$ series, which are
solvable analytically. Moreover, each correlation function gets
a contribution from only one genus. This is an indication that
these models may be identified with those of topological gravity.

\appendix{(1,q) Models}

The $(1,q)$ models are specified by the choice,
$$P = R_+ \equiv D.	\eq{Pop}$$
This enables one to solve the string equation explicitly for
the $x$ dependence,
$$Q = D^q + x.	\eq{Qopnew}$$
As discussed in detail in [\putref{mont,rivl2}], this
allows one to write exact formulae for the $R^i$ 
$$\eqalign{R^i = & D^i + {{i}\over{q}} x D^{i-q} + {{i(i-q)}\over{2q}}
D^{i-q-1} \cr & + {{i(i-q)}\over{2q^2}} x^2 D^{i-2q} + {{i(i-q)(i-2q)}\over
{2q^2}} x D^{i-2q-1} \cr & + {{i(i-q)(i-2q)(3i-5q-4)}\over{24 q^2}} D^{i-2q-2} +
{{i(i-q)(i-2q)}\over{6q^3}} x^3 D^{i-3q} + ...\cr}\eq{Ri}
$$
and for various correlation
functions. We refer the reader to those papers for details.
We showed that one may associate a `ghost' number to
each operator,
$$\gh(\cp_i) = q+1 - i,  \eq{ghostnumber}$$
and that the correlation functions obey certain selection rules. The first one
states that the correlation function $\bra\cp_{i_1}\ldots\cp_{i_n}\ket$ will
vanish unless,
$$\sum_{j=1}^{n}\gh(\cp_{i_j}) \le 2(1+q). \eq{selection1}$$
This is true even for $x\ne 0$. The other selection rule which is true
only for $x=0$ (\ie, in the topological limit) is,
$$\sum_{j=1}^n \dim(\cp_{i_j}) = \sum_{j=1}^n i_j = 0 \pmod{q+1}.
\eq{selection2}$$
Combining the two together one derives the ghost number conservation law,
$$\sum_{j=1}^n \gh(\cp_{i_j}) = 2(1+q)(1-g),  \eq{conservation}$$
where $g$ is a positive integer that is identified with the genus.

Now, as promised earlier, we will discuss the integration constants.
As explained in our previous paper, the reason $Q$ is equal to $D^q+x$
is because one can associate a scaling dimension to the operators.
The result is that the function $u_0=x$ has dimension $q$. Derivatives
contribute dimension $1$ and thus constants have dimension $0\pmod{q+1}$.
Thus, when integrating an equation, one can add constants only when
the dimension of the correlation function is $0 \pmod{q+1}$.
Thus, for example, in the transition from (\putlab{p1p1pm}) to
(\putlab{p1pm}) the constant $\alpha_m$ may not be zero only for
$m=q \pmod{q+1}$! In general a `constant' means only that it's a constant
with respect to $x$ (or $t_1$). So, when taking flows with respect to
the other $t_i$'s one has to be careful and try to compute those constants.
Only correlation functions that have two punctures (\ie, two insertions
of $\cp_1$) do not involve integration. 

\appendix{Extension of (1,q) Models and the Metric}

Since the operators $R^i$ exist for negative $i$\footnote[*]{The explicit
formula in reference [\putref{mont}] is valid for negative powers.},
one may extend the range of fields $\cp_i$ to include
negative indices. Doing this, one easily sees that since
$\ord R^i = i $ then $R^i_+ = 0$ for $i<0$. Thus, the KdV flows
with respect to these operators are trivial. That is
$${{\partial R^i}\over{\partial t_{j}}}=\lb R^j_+, R^i\rb = 0 \quad
{\rm for} \quad j\le0. \eq{negflows}$$
This yields the not too surprising result that correlations functions
that contain these fields vanish. For example,
$$\bra\cp_1\cp_1\cp_i\cp_j\ket = 0\quad for \quad i<0.  \eq{zero}$$

However, there are some correlation functions that do not vanish. For example,
$$\bra \cp_1 \cp_m \ket = \Res_{-1} R^m \eq{tt}$$
does not vanish for $m=-1$. Actually one can evaluate it exactly and get
$$\bra \cp_1\cp_{-1}\ket = 1. \eq{norm1}$$
It is clear that correlation functions that involve `flowing' of
$\bra\cp_1\cp_m\ket$ with a negative flow will vanish (such as (\putlab{zero})),
but some that involve direct integration might be non zero.

The general two-point correlation function on the sphere,
$\bra\cp_i\cp_j\ket_0$,
will vanish unless $i+j=0$. This is a consequence of the ghost number
conservation rule (\putlab{conservation}). In the regular theory this
correlation function vanishes always because there are no negative indexed
fields. However, after adding the new fields some 
two-point functions may be non zero,
and we now wish to compute them.
 That will enable us to define a metric in a way
similar to regular (\ie, without gravity) topological field theory.
We conjecture that the result is
$$\bra\cp_i \cp_j\ket = |i|\delta_{i+j,0}. \eq{metric}$$
We will prove it for low values of $i$ and $j$.

Starting with $\bra \cp_1 \cp_i \ket$ we flow with $t_j$ and then integrate
out the $\cp_1$. For $j<q$ one has
$$\eqalign{\bra\cp_1\cp_i\cp_j\ket= & \Res_{-1}\lb D^j, R^i\rb \cr
= & \sum_{k=0}^{j-1}{{j}\choose{k}} \Res_{-1}\left(R^i\right)^{(j-k)} D^k \cr
= & \sum_{k=0}^{j-1}{{j}\choose{k}}\left( \Res_{-1-k}R^i\right)^{(j-k)}.
\cr} \eq{st1}$$
Integrating, we get
$$\bra\cp_i\cp_j\ket =
\sum_{k=0}^{j-1}{{j}\choose{k}}\left( \Res_{-1-k}R^i\right)^{(j-k-1)},
\eq{met1}$$
and substituting $i=-j$,
$$\bra\cp_{-j}\cp_j\ket = j  \quad {\rm for} \quad q>j>0. \eq{met2}$$

To compute the metric for non-primary fields we need to introduce a trick.
This involves the computation of derivatives of $R^i$. It is based on the
observation that (note that this is only true for the $(1,q$ models),
$$ Q' = \left( R^q \right)' = 1.  \eq{derQ}$$
Thus, we get that
$$R' = {{1}\over{q}} R^{1-q}, \eq{derR}$$
and, generalizing, we get\footnote[*]{This identity can be used to compute the
partition function and various one-point functions (by integrating $\cp_1$)
and also to prove the `puncture equation' directly from KdV gravity. We will
do it later.}
$$\left( R^i\right)^{(j)} = {{i(i-q)(i-2q)\cdots(i-(j-1)q)}\over{q^j}}
R^{i-j q}.  \eq{derRi}$$
Another useful identity is
$$Q\left( R^i\right)^{(j)} = {{i-(j-1)q}\over{i+q}}\left( R^{i+q}\right)^{(j)}.
\eq{QderRi}$$

So, for $q>j>0$ we can compute
$$\eqalign{\bra\cp_1\cp_i\cp_{j+q}\ket = & \Res_{-1}\lb R^{q+j},R^i\rb \cr
= & \Res_{-1}\lb D^{q+j} + {{q+j}\over{q}} x D^j +
{{j(q+j)}\over{2q}} D^{j-1}, R^i\rb\cr
= & \Res_{-1} \lb {{q+j}\over{q}} Q D^j - {{j}\over{q}} D^{q+j} +
{{j(q+j)}\over{2q}} D^{j-1}, R^i\rb\cr
= & \Res_{-1}\Bigl( {{q+j}\over{q}}\sum_{k=0}^{j-1}{j\choose k} Q
\left( R^i\right)^{(j-k)} D^k - {{j}\over{q}} \sum_{k=0}^{q+j-1}
{{q+j}\choose k} \left( R^i\right)^{(q+j-k)} D^k \cr
& + {{j(q+j)}\over{2q}} \sum_{k=0}^{j-2}
{{j-1}\choose k} \left( R^i\right)^{(j-1-k)} D^k \Bigr).\cr
}\eq{st2}$$
Thus,
$$\eqalign{\bra\cp_i\cp_{j+q}\ket = & {{q+j}\over{q}}\sum_{k=0}^{j-1}
{j\choose k} {{i-(j-k-1)q}\over{i+q}}\left(\Res_{-1-k}
R^{i+q}\right)^{(j-k-1)} \cr
& - {{j}\over{q}} \sum_{k=0}^{q+j-1} {{q+j}\choose k} \left(\Res_{-1-k}
R^i\right)^{(q+j-k-1)} \cr
& + {{j(q+j)}\over{2q}} \sum_{k=0}^{j-2} {{j-1}\choose k} \left(\Res_{-1-k}
R^i\right)^{(j-2-k)},\cr
}\eq{met3}$$
and setting $j=-i-q$ we get
$$\bra\cp_{-i-q}\cp_{i+q}\ket = q+i\quad{\rm for}\quad q>i>0.\eq{met4}$$
We computed the metric (\putlab{metric}) for $2q<i<3q$ as well, and the result
still holds. The computation is not illuminating and will be omitted.
Through induction it may now be possible to prove (\putlab{metric}) for all $i$.

By an alternative method, we can compute the metric for any $i$. This method
as not as convincing as the previous one because it involves analytic 
continuation of integers to real numbers. The starting point is the following
correlation functions derived from the KdV approach.
$$\bra\cp_{qi + \alpha} \cp_{qj + q-\alpha} \cp_1^{i+j+1}\ket_0 = 
q {{i+j}\choose{i}} {{\Gamma(i+1+{\alpha\over q})}\over 
{\Gamma(1+{\alpha\over q})}} {{\Gamma(j+1+{{q-\alpha}\over q})}
\over{\Gamma(1+{{q-\alpha}\over q})}}. \eq{pipjp1}$$
This equation is true for all non-negative (integer) $i$ and $j$. By
analytically continuing the result to $i+j+1=0$ one can show, after
some algebra, that
$$\bra\cp_{qi+\alpha}\cp_{-qi-\alpha}\ket = q i +\alpha.  \eq{METRIC}$$
\appendix{Integration and the Puncture Equation}

Using equation (\putlab{derRi}) we can integrate the two-point correlation
function (\putlab{tt}) and get the one-point function. By using the identity
$R^m = {q/(q+m)} (R^{m+q})'$, we get
$$\bra \cp_m \ket = {{q}\over{q+m}} \Res_{-1} R^{q+m}. \eq{onepoint}$$
For example, the dilaton expectation value is
$$\bra \cp_{q+1}\ket =  {{q^2-1}\over{24q}}  \eq{dilaton}$$

Similarly, we can integrate (\putlab{p1p1}) twice and get the partition
function
$$\bra 1 \ket = {{q^2}\over {(q+1)(2q+1)}} \Res_{-1} R^{2q+1}
={{q-1}\over{24}}.  \eq{part}$$
One should note that one can {\it not} flow these equations because we
have set all the flow variables to zero by using the explicit form
of $Q$ (except for $x$, of course).

To get the puncture equation we use a variant of (\putlab{derRi}):
$$\left( R^i \right)^{(j)} = {{i}\over{q}} \left( R^{i-q} \right)^{(j-1)}.
\eq{dervar}$$
Thus, using also the KdV flows, one can show that
$$\bra\cp_1^{k+1}\cp_{i_1}\cp_{i_2}\cdots\cp_{i_n}\ket =
\sum_{j=1}^n {{i_j}\over{q}}\bra\cp_1^k\cp_{i_1}
\cdots\cp_{i_{j-1}}\cp_{i_j-q}\cp_{i_{j+1}}\cdots\cp_{i_n}\ket \eq{puneq}$$
which is the puncture equation. It was the the first example of a recursion
relation in topological gravity and, as we have seen, can be proven
directly from the KdV.

\appendix{The $W_3$ Recursion Relations}

In this appendix we will review the $W_3$ constraints and present
the corresponding recursion relations on the sphere. These results 
should be compared with those of Section 4. 

Let us consider $(1,q)$ models. For correlation functions involving
the $n$th descendant of the puncture operator as the first 
operator\footnote[*]{The notation is defined in the appendices.},
$$\bra \cp_{nq+1}\prod_{i\in S} \cp_i \ket,$$
the Virasoro constraints can be used to derive recursion 
relations\ref{dijk3,fuku}. Since
we actually have $q-1$ primary fields, more information is needed to
completely solve the models.

The $W_3$ constraints can be used to derive
recursion relations for correlation functions involving descendants of
the second primary field as the first operator,
$$\bra \cp_{nq+2}\prod_{i\in S} \cp_i \ket.$$
For the $(1,3)$ model K.~Li\ref{li} has shown 
that the $W_3$ constraints are given by $W_m\tau=0$ with  
$m\geq -2$ and
$$\eqalign{
W_m=& {16\over 9}{\partial \over \partial t_{m+2,2}} -{8\over 3}
\sum_{r-p=m+1}(p+{1\over 3})t_{p,1}{\partial \over \partial t_r,2}
-{4\lambda^2 \over 3}\sum_{r+q=m}{\partial^2 \over \partial t_{r,2}
\partial t_{q,2}} \cr
&+\sum_{p+q-r=-m}(p+{1\over 3})(q+{1\over3})t_{p,1}t_{q,1}{\partial
\over \partial t_{r,2}}\cr
&+\sum_{p+q-r=-m-1}(p+{2\over 3})(q+{2\over 3})t_{p,2}t_{q,2}{\partial
\over \partial t_{r,1}}\cr
&+\lambda^2\lbrace\sum_{p-q-r=-m+1}(p+{1\over 3})t_{p,1}{\partial^2
\over\partial t_{q,2} \partial t_{r,2}}\cr
&+\sum_{p-q-r=-m}(p+{2\over 3})t_{p,2}{\partial^2\over \partial t_{q,1}
\partial t_{r,1}}\rbrace\cr
&+{\lambda^4\over 3}\lbrace\sum_{p+q+r=m-1}{\partial^3\over\partial t_{p,1}
\partial t_{q,1} \partial t_{r,1}}+\sum_{p+q+r=m-2}{\partial^3\over
\partial t_{p,2} \partial t_{q,2} \partial t_{r,2}}\rbrace \cr
&+{1\over 27}\lambda^{-2}((8t_{0,2}^3-4t_{0,1}^2)\delta_{m,-2} +
t_{0,1}^3\delta_{m,-1}),\cr}\eq{wgen}$$
where $\lambda$ is the string coupling constant. This was derived by requiring
the consistency of the commutation relations of the $W_3$ algebra.
From the above one can now read off the recursion relations. In particular,
we are interested in the recursion relations on the sphere.

Let us consider the following set of correlation functions:
$$\bra\cp_{3i+2}\cp_1^{i+2}\ket_0=\prod_{j=0}^i (j+{2\over 3}).\eq{exactkdv}$$
This result can be obtained from either the KdV\ref{mont} or the Virasoro
constraints. From the $W_3$ constraint, (\putlab{wgen}), 
the recursion relation for the above correlation functions is
$$\eqalign{&
\bra\cp_{3i+2}\cp_1^{i+2}\ket_0=\cr
&+a_0(i+2){1\over 3}\bra\cp_{3(i-1)+2}
\cp_1^{i+1}\ket_0 
+a_1{(i+2)(i+1)\over 2}{1\over 9}\bra\cp_{3(i-2)+2}\cp_1^i\ket_0 \cr
&+a_2\sum_{a=2}^{i}{i+2\choose a}\bra\cp_{3(a-2)+2}\cp_1^a\ket_0
\bra\cp_{3(i-a)+2}\cp_1^{i+2-a}\ket_0\cr
&+a_3\sum_{a=2}^{i}\sum_{b=2}^{i-a}{i+2\choose a}{i+2-a\choose b}
\bra\cp_{3(a-2)+2}\cp_1^a\ket_0\bra\cp_{3(b-2)+2}\cp_1^b\ket_0
\bra\cp_{3(i-a-b)+2}\cp_1^{i+2-a-b}\ket_0 \cr
&+a_4\sum_{a=3}^{i}{i+2\choose a}a{1\over 3}\bra\cp_{3(a-3)+2}\cp_1^{a-1}
\ket_0\bra\cp_{3(i-a)+2}\cp_1^{i+2-a}\ket_0 +a_5 \delta_{i,1}
 +a_6\delta_{i,2}.\cr}
\eq{wrecurs}$$
The different terms in this equation correspond to double contact
terms, triple contact
terms, a single surface degeneration, a double surface degeneration,
a surface degeneration plus a contact term, and the $\delta$
functions in equation (\putlab{wgen}), respectively. With 
a normalization consistent with the
KdV results (as presented in the appendices):
$$\eqalign{a_0 & =2,\qquad a_1 = -2,\qquad a_2= {{1}\over{3}},\qquad 
a_3=-{{1}\over{27}}\cr a_4 & = -{{1}\over{3}},\qquad a_5=-{{2}\over{9}}
\ \ \ {\rm and}\ \ \ a_6 = {{2}\over{3}}. \cr} \eq{avalues}$$

Equation (\putlab{wrecurs}) is not enlightening and tedious to extend
to general $(1,q)$ models; in particular, the $\delta$ functions are somewhat
difficult to explain. This should be contrasted with equation 
(\putlab{ex1cont}).

\newpage

\lefttext{\bf\ References}
\medskip

\begin{putreferences}

\def\ref#1{\markup{[\putref{#1}]}}
\def\nuc#1{{\it Nucl. Phys.}, {\bf B#1}}
\def\pl#1{{\it Phys. Lett.}, {\bf #1B}}
\def\cmp#1{{\it Comm. Math. Phys.}, {\bf #1}}

\def\prl#1{{\it Phys. Rev. Lett.}, {\bf #1}}
\def\mpl#1{{\it Mod. Phys. Lett.}, {\bf A#1}}

\reference{bess}{%
D.~Bessis, \cmp{69} (1979) 147.}

\reference{bess2}{%
D.~Bessis, C.~Itzykson and J.-B.~Zuber, {\it Adv. App. Math.}, (1980) 109.}

\reference{brez}{%
E.~Brezin, C.~Itzykson, G.~Parisi and J.-B.~Zuber, \cmp{59} (1978) 23.}

\reference{brez2}{%
E.~Brezin and V.~A.~Kazakov, \pl{236} (1990) 144.}

\reference{davi}{%
F.~David, \nuc{257} [FS14] (1985) 433.}

\reference{davi2}{%
F.~David, \mpl{3} (1988) 1651.}

\reference{dijk}{%
R.~Dijkgraaf and E.~Verlinde, {\sl Annecy Field Theory} 1988:87.}

\reference{dijk2}{%
R.~Dijkgraaf and E.~Witten, \nuc{342} (1990) 486.}

\reference{dijk3}{%
R.~Dijkgraaf, E.~Verlinde and H.~Verlinde, \nuc{348} (1991) 435.}

\reference{dijk4}{%
R.~Dijkgraaf, E.~Verlinde and H.~Verlinde, \nuc{352} (1991) 59.}

\reference{dist}{%
J.~Distler, \nuc{342} (1990) 523.}

\reference{dist2}{%
J.~Distler and H.~Kawai, \nuc{321} (1989) 509.}

\reference{doug2}{%
M.~R.~Douglas and S.~Shenker, \nuc{335} (1990) 653.}

\reference{doug3}{%
M.~R.~Douglas, \pl{238} (1990) 176.}

\reference{eguc}{%
T.~Eguchi and S.-K.~Yang, \mpl{5} (1990) 1693.}

\reference{fuku}{%
M.~Fukuma, H.~Kawai and R.~Nakayama,
 {\it Int. J. Mod. Phys.} {\bf A6} (1991) 1385.}

\reference{goer}{%
J.~Goeree, \nuc{358} (1991) 737.}

\reference{gros}{%
D.~Gross and A.~A.~Migdal,  \prl{64} (1990) 127; \nuc{340} (1990) 333.}

\reference{itzy}{%
C.~Itzykson, \nuc{5} (Proc. supl., 1988) 150.}

\reference{kaza2}{%
V.~A.~Kazakov, \pl{50} (1985) 282.}

\reference{kaza3}{%
V.~A.~Kazakov, I.~K.~Kostov and A.~A.~Migdal, \pl{57} (1985) 295.}

\reference{kaza4}{%
V.~A.~Kazakov A.~A.~Migdal, \nuc{311} (1988/89) 171.}

\reference{li}{%
K. Li, \nuc{354} (1991) 711; \nuc{354} (1991) 725.}

\reference{mont}{%
D.~Montano and  G.~Rivlis, \nuc{360} (1991) 
524.}

\reference{rivl2}{%
G.~Rivlis, {\it Two Topics in 2D Quantum Field Theory}, Ph.D. Thesis, 
Caltech, 1991.}

\reference{verl}{%
E.~Verlinde, \nuc{300} (1988) 360.}

\reference{verl2}{%
E.~Verlinde and H.~Verlinde, \nuc{348} (1991) 457.}

\reference{witt}{%
E.~Witten, \cmp{92} (1984) 455.}

\reference{witt2}{%
E.~Witten, \nuc{340} (1990) 281.}

\end{putreferences}

\bye